\documentclass[12pt,a4]{article}

\usepackage[top=2.5cm, bottom=2.5cm, left=2.5cm, right=2.5cm]{geometry}

\usepackage{alltt}
\usepackage{graphicx}
\usepackage{color}
\usepackage{nicefrac}
\usepackage{subfigure}
\usepackage{amsmath}
\usepackage{amssymb}
\usepackage{float}
\usepackage{framed}
\usepackage{stmaryrd}
\usepackage{tikz}
\usepackage{tkz-graph}
\usepackage{lmodern}
\usepackage{fancyvrb, relsize}
\usepackage{listings}
\usepackage{subfigure}
\usepackage{float}
\usepackage{multirow}
\usepackage{paralist}
\usepackage{alltt}
\usepackage{array}
\usepackage{xspace}
\usepackage{relsize}
\usepackage[longend,linesnumbered,ruled]{algorithm2e}

\newcommand{\failure}{{\mathsf{failure}}}
\newcommand{\fail}{\otimes}
\newcommand{\succeed}{\checkmark}

\newcommand{\hide}[1]{}

\newcommand{\size}[1]{{\mid \! {#1}\! \mid}}

\newtheorem{assumption}{Assumption}

\begin{document}

\begin{center}
{\Large \bf Optimizing Governed Blockchains\\
\smallskip
for Financial Process Authentications}\\

\medskip
{\small Leif-Nissen Lundb{\ae}k, Andrea Callia D'Iddio, and Michael Huth\\
Department of Computing, Imperial College London\\
London, SW7 2AZ, United Kingdom\\
$\{$leif.lundbaek, a.callia-diddio14, m.huth$\}$@imperial.ac.uk}
\end{center}

\date{\today}

\begin{abstract}
We propose the formal study of governed blockchains that are owned and controlled by organizations and that neither create cryptocurrencies nor provide any incentives to solvers of cryptographic puzzles. We view such approaches as frameworks in which system parts, such as the cryptographic puzzle, may be instantiated with different technology. Owners of such a blockchain procure puzzle solvers as resources they control, and use a mathematical model to compute optimal parameters for the cryptographic puzzle mechanism or other parts of the blockchain. We illustrate this approach with a use case in which blockchains record hashes of financial process transactions to increase their trustworthiness and that of their audits.
For Proof of Work as cryptographic puzzle, we develop a detailed mathematical model to derive MINLP optimization problems for computing optimal Proof of Work configuration parameters that trade off potentially conflicting aspects such as availability, resiliency, security, and cost in this governed setting. We demonstrate the utility of such a \emph{mining calculus} by applying it on some instances of this problem. This experimental validation is strengthened by statistical experiments that confirm the validity of random variables used in formulating our mathematical model.
We hope that our work may facilitate the creation of \emph{domain-specific} blockchains for a wide range of applications such as trustworthy information in Internet of Things systems and bespoke improvements of legacy financial services.
\end{abstract}

\section{Introduction}
\label{section:introduction}
There is little doubt that modern accounting systems have benefitted, ever since the advent of commercial computing machines, from the digitization of the processing and recording of financial transactions. The automated processing of payroll information in the 1950ies was perhaps one of the earliest examples of such benefits: IBM introduced its \emph{702 Data Processing System} for businesses in 1953. And the use of RFID technology or smart phones for contactless payment of small items such as coffees is a more recent example thereof.

It is then striking that the mechanisms used for managing the integrity of accounts are, in essence, those developed at least a thousand years ago. What we call the modern
\emph{double-entry bookkeeping} was already used by Florentine merchants in the 13th century, for example. Without going into great detail, the key idea is in simplified terms that each account has an associated \emph{dual} account and that each credit in one account is recorded as a debit in that dual account. This allows for the formulation and verification of an important \emph{financial invariant}: no matter how complex financial transactions may be, or how many transactions may occur, it must always be the case that over the totality of accounts
\begin{quote}
``Assets equal liabilities plus capital.''
\end{quote}

Modern realizations of this method may enrich account entries with time stamps and other contextual data so that the flow of assets can be better understood, for example to support an audit. The above invariant may be quite simple to verify, and its verification may give us reassurance that every debit has an associated credit. But it does not prevent the recording of transactions that may be unauthorized, fraudulent, or that may be incorrect due to human error.
For example, transaction records within accounting books may be manipulated to commit fraud whilst these manipulations still satisfy the above invariant.

One may say that processing of transactions is governed by a form of \emph{legal code} that is informed by policy on fraud prevention and detection, regulation, compliance, risk, and so forth. But the enforcement of such legal code within the \emph{technical code} that operationalizes modern financial processes has been difficult at best, and too costly or impossible at worst.

Digitized financial processes can, of course, utilize cryptographic primitives to help with narrowing this gap between legal and technical code: digital signatures can be associated to transactions (for example\ embedded within transaction objects), and commitment schemes can be used to realize consistent distributed storage whose consistency is resilient to adversarial manipulation; see for example\ the discussion of Byzantine Agreement Protocols in \cite{wattenhofer16}. But the advent of de-centralized, \emph{eventual consistency} storage protocols, as pioneered in the cryptocurrency Bitcoin \cite{nakamoto08}, opened up a new way of thinking about the processing of financial transactions, even of creating and managing a currency as a unit of account. There is little doubt that cryptocurrencies are one of the most important innovations \cite{ali14,narayanan16}, along with the invention and introduction of central banks, in financial services since the advent of the double-entry bookkeeping.

In Bitcoin, transactions are grouped into blocks, and blocks are recorded and linked in a chain of blocks~--~the blockchain. Currency units are created through the solution of a cryptographic mining puzzle, a process in which network nodes (called miners) compete in determining the next block to be added to the chain and where the winner will become the owner of the currency created in that new block. These solutions causally link this new block to the last one, using cryptographic hash functions, creating thus the resiliency of this chain against manipulation.  These acts of creating currency are treated as (special) transactions and their outputs are associated with their owners in a (pseudo)anonymous manner~--~using public-key cryptography. The system dynamics is adjusted so that, on average, a new block is added to the blockchain about every 10 minutes.

In Bitcoin, transactions are now technical code. Without going into the technical details, a transaction can be seen as a program that has a number of inputs and a number of outputs, each representing a unit of currency associated with some owner. That program also contains a \emph{proof} that the program is authorized to rewire inputs to outputs in this manner. For example, this proof may be a cryptographic demonstration that the program \emph{owns} all inputs. Transactions enjoy an important invariant: the sum of currency units of all transaction outputs cannot be larger than the sum of currency units of all its inputs~--~and any positive difference becomes a credit for the miner that created the block within which the transaction is found.

The Bitcoin network has many nodes that contain a full copy of the blockchain. More accurately, each node contains a tree of possible chains and identifies one of those chains deterministically as the \emph{real} blockchain. The propagation of possible winning blocks and consensus protocols between these nodes ensure that nodes eventually agree on which of the chains they each store is the \emph{real} chain. This does have its problems. For example, miners may collude to gain more control of mining competitions and so may force all network nodes to accept an alternative blockchain as the real one~--~allowing them to rewrite the transaction history. Another problem is that nodes need to keep a record of which outputs of transactions recorded on the \emph{real} blockchain have not been spent yet in new transactions on the blockchain; and the above consensus mechanisms for \emph{eventual consistency} don't give hard guarantees, meaning that transaction outputs may be double spent and that transactions in a blockchain may only be trusted if their block has been linked with a certain number of blocks added to the chain subsequently. The incentive mechanism of Bitcoin also led to the creation of puzzle-solving pools of miners and a complex and risky dynamics between such pools, the development of hardware for solving such puzzles and other factors. Understanding such dynamics in predictable ways is one research challenge for cryptocurrencies \cite{DBLP:conf/sp/BonneauMCNKF15}.

Another risk resides in the verification of transactions, done by a run-time system that executes a verification script. Whilst the designer(s) of Bitcoin deliberately chose a simple scripting language that makes its execution more secure, early implementations of that system still had security vulnerabilities. There is also a tradeoff here between security and expressiveness of technical code. Other cryptocurrencies (see Chapter~10 in \cite{DBLP:conf/sp/BonneauMCNKF15}) seek to program more complex transactions (so called \emph{smart contracts}) to create new financial services, for example. But these require scripting languages that are Turing complete and are therefore much more at risk of security attacks~--~see the DAO hack of Ethereum and the way in which this was dealt with as a good example thereof \cite{castillo16}.

The risks of cryptocurrencies discussed above suggest that there is
value in also exploring alternative approaches, in which technical code for transactions is more centralized, governed or controlled by parties with specific interests or duties. For example, RSCoin \cite{DBLP:journals/corr/DanezisM15} is proposed as a cryptocurrency in which central banks control monetary supply and where a number of distributed authorities prevent double-spending attacks. This addresses or mitigates risks of de-centralized, open cryptocurrencies but does support a more conventional model of currency control. There is also the question of whether a central bank would want to risk running any such system to scale up a currency nationally, given that such technical code may be subject to security vulnerabilities in configuration, implementation or lifecycle management. But such risks may be manageable. The oldest central bank, Sweden's Riksbank, for example, is actively considering whether or not to introduce a national cryptocurrency \emph{ekrona} within the next two years \cite{Milne16}. 

In this paper, we investigate how governed, closed blockchains can be designed so that they can support the resilient, distributed, and \emph{trustworthy} storage of authentication of transactions within
conventional financial processes. Such governed systems with restricted access give us better control on balancing the use of energy for puzzle solving with the security of the Proof of Work algorithm when compared with open systems that rely on Proof of Work, such as Bitcoin.
Specifically, we propose that transactions (in the sense of Bitcoin) within blocks are hashes of transactions (in the sense of conventional financial processes). We then define mathematical models that describe the design space of such a blockchain in terms of the cryptographic puzzle used~--~in this paper Proof of Work, in terms of expected availability, resiliency, security, and cost, and in terms that reflect that the system is centrally governed.

We stress that our approach is also consistent with transactions within blockchains that encode transaction history, which we don't consider in the use case of this paper. We believe that our approach has potential. It may, for example, allow designers to minimize the need for consensus mechanism by guaranteeing that puzzles, with very large probability, have a unique winner within a certain period of time whilst still maintaining sufficient system resiliency and security; and this could inform the design of bespoke consensus protocols.

\paragraph{Outline of paper.} In Section~\ref{section:usecase} we present our use case. Our mathematical model for Proof of Work for our setting is subject of Section~\ref{section:probabilisticmodel}. The derivation of optimization problems for these mathematical models is done in Section~\ref{section:optimization}. An algorithm for solving such optimization problems, experimental results, and a statistical validation of our model are reported in Section~\ref{section:experiments}, and the paper concludes in Section~\ref{section:conclusion}.

\section{Use case}
\label{section:usecase}
The use case we consider is one of a financial process that creates financial transactions. We would like to enhance the trustworthiness of this process through a blockchain that records hash-based authentications of transactions, as seen in Figure~\ref{fig:architecture}, where the interaction between the legacy process and the blockchain is conceptually simple~--~and consistent with the use of double-entry bookkeeping if desired. Our assumption is that the event streams of such transactions are not linearizable and so we cannot rely on techniques such as hash chains~\cite{DBLP:reference/crypt/Horne11} to obtain immutability of transactions. Moreover, a hash chain could be recomputed with little effort by an attacker with partial control of the system. A blockchain is much more resilient to such an attack as it takes considerable effort to solve the number of cryptographic puzzles needed for rewriting parts or all of a blockchain.

Our data model represents a transaction as a string $input$ that can be authenticated with a hash $hash(input)$. String $input$ may be a serialization of a transaction object that contains relevant information such as a time stamp of the transaction, a digital signature of the core transaction data and so forth. The trustworthiness of transaction $input$ is represented outside of the blockchain by the triple
\begin{equation}
\label{equ:triple}
(input,hash(input), location)
\end{equation}

\noindent where $location$ is either the block height ($\geq 0$) of a block $b$ in the blockchain such that $hash(input)$ occurs in block $b$ or $location$ is NULL, indicating that the transaction is not yet confirmed on the blockchain.

The hashes $hash(input)$ of transactions that still need to be confirmed are propagated on the blockchain network, where they are picked up by miners and integrated into blocks for Proof of Work. We assume a suitable mechanism by which nodes that manage legacy accounts learn the blockheights of their transactions that have been successfully added to the blockchain. For example, such nodes may have a full copy of the blockchain and so update $location$ values in their corresponding accounts if the hash of the corresponding transaction occurs in a block that was just added.

A transaction is unverified if its $location$ value is NULL or if its hash does not equal the one stored externally as in~(\ref{equ:triple}); it is trustworthy if $0\leq location$ and $location + k\leq currentBlockH\! eight$ where $k\geq 0$ is a suitable constant and $currentBlockH\! eight$ denotes the number of blocks added to the blockchain so far. The value of $k$ may be a function of how fast blocks are added to the chain on average, to ensure sufficient resiliency of trustworthiness.
 An auditor could then inspect any transaction by examining its triple stored as in~(\ref{equ:triple}). If $location$ equals NULL or if $location + k > currentBlockH\! eight$, the transaction is considered neither valid nor trustworthy by the auditor.
Otherwise, we have $0\leq location$ and $location + k \leq currentBlockH\! eight$ and the auditor uses the Merkle tree hash in block $location$ (or whatever efficient mechanism for membership tests is provided in the blockchain data structure) to verify that $hash(input)$ is in the block of height $b$. If that is the case, the auditor considers the transaction to be verified; otherwise, the auditor considers the transaction not to be trustworthy.

Note that this use case does not require transaction scripts to be stored, nor any run-time system for verifying such transactions. But the modelling approach we present in this paper is consistent with use cases that have such script generation and verification support.

\subsection{System Architecture}
\begin{figure}[h]
\centering
\includegraphics[scale=0.64]{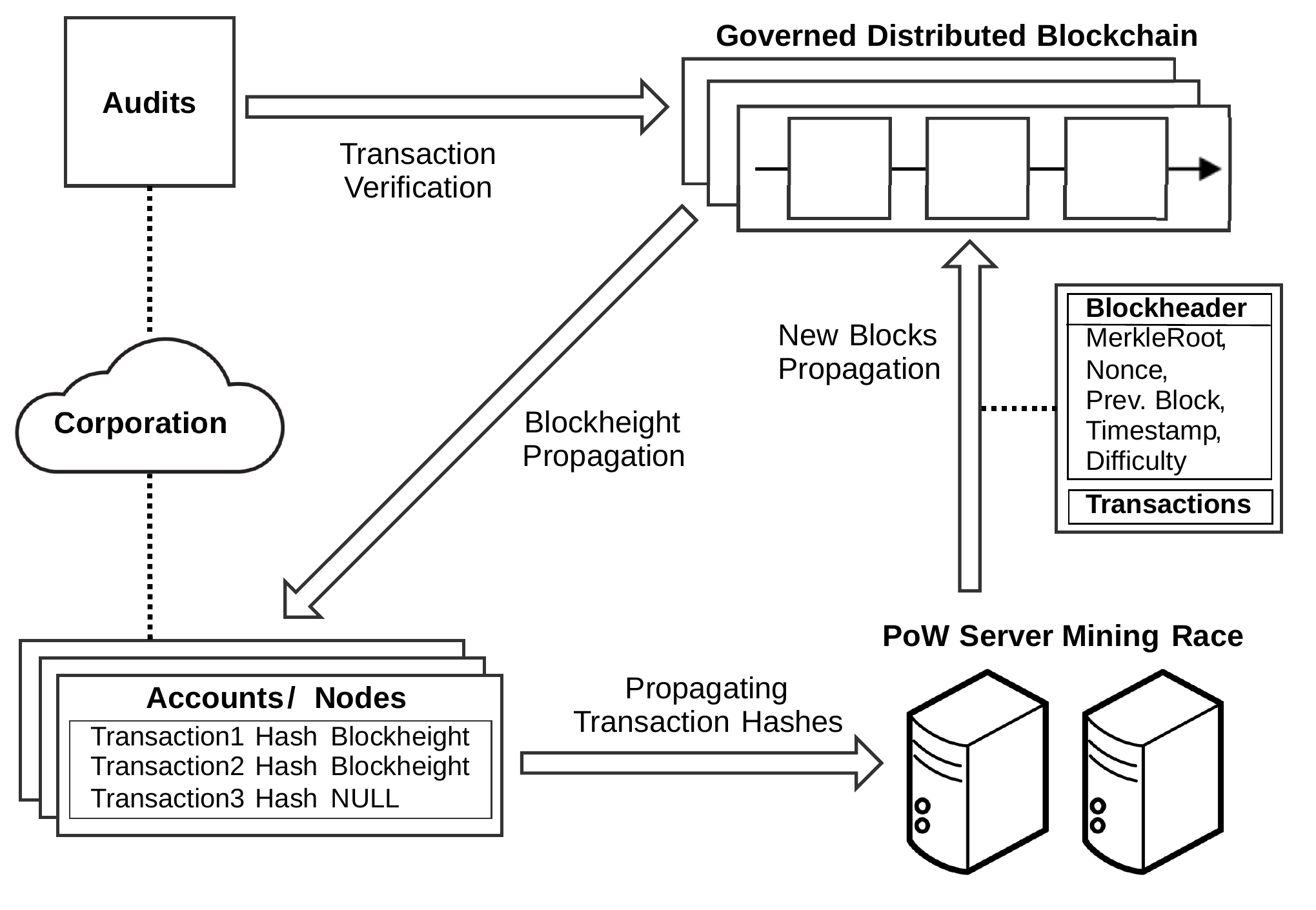}
\vspace{-0.4cm}
\caption{Governed blockchain for financial process authentications. In our use case, {\bf Transactions} within blocks on the left are mere hashes of transactions listed in {\bf Accounts / Nodes} on the left
\label{fig:architecture}}
\end{figure}

A system architecture that could support such a use case is shown in Figure~\ref{fig:architecture}. Unverified transactions have their hashes propagated on the network. Miners pick up those hashes and integrate them into blocks for Proof of Work. We abstract away how miners manage their pools of hashes and how Proof of Work blocks are propagated and added to the blockchain; this gives us flexibility in the use of blockchain technology.
Once blocks are added to the blockchain, blockheights are propagated to the legacy account. As mentioned above, these accounts could have full copies of the blockchain and thus implement their own update mechanisms for value $location$ in triples stored as in~(\ref{equ:triple}).

Auditors would interface with both accounts and the blockchain to verify, in a trustworthy manner, the authenticity of transactions. Any transaction that is not verified as discussed above would be flagged up in this audit. Any pre-existing audit process~--~which may focus on compliance, regulations and other aspects~--~is consistent with such \emph{trustworthiness checking}; and the trustworthiness of the pre-existing audit process would be increased as it would refuse to certify any financial transaction histories that involved a transaction that is not authenticated on the blockchain.

\subsection{Discussion}
The approach we advocate here is pretty flexible. It seems consistent with consensus mechanisms as used in Bitcoin but it may also support 2-phase commitment schemes as proposed in \cite{DBLP:journals/corr/DanezisM15}. Our system architecture allows for full nodes to be associated with accounts, sets of accounts or corporate boundaries. Our blockchain does not create any currency, and so there is no inherent incentive to mine. But there is an incentive for the owners of this blockchain to allocate mining resources in a manner that establishes
trustworthiness of transactions as recorded in this blockchain. We think that the elimination of incentives and their game-theoretic implications are a benefit, as are the relatively simple ways of propagating trust through hashes of transactions. Such a blockchain may also be consulted by legacy systems to inform the authorization of further financial transactions.

Our blockchain does not spend any funds and so has no problem of \emph{double spending}, and double spending in the legacy system would be detectable with existing mechanisms such as audits. Our approach does allow for \emph{double authentication} though: a transaction hash may occur more than once in a blockchain, be it in the same block or in different blocks. We deem this to be unproblematic as audits would only need to establish \emph{some}, sufficiently old, authentication of the transaction in the blockchain to establish its trustworthiness~--~noting that hash-based authentication is deterministic.

One expectation is that blocks would only be added to the blockchain if they were signed by one of the miners that is resourced for this Proof of Work service. This requires that the public keys of such miners are securely stored and available within the system. Over time, some of these miners may be removed from such a list (e.g.\ decommissioned) and new ones may be added (e.g.\ system upgrade). Change in the number of miners is then a configuration change, they would also change the local nonce space for miners in the mathematical model we will develop in this paper. But such change is locally computable from a new tuple value $(r,s,d)$.

Let us now discuss an attack model for this use case shown in Figure~\ref{fig:architecture}. Other attack models may apply to such a use case as well, and our mathematical model and analysis approach is not tied to the particular attack model. There are various actors within that system: miners, nodes that run full copies of the blockchain, nodes that propagate transaction hashes to miners, auditors (which may be insiders or external agents with limited inside access), insiders such as accountants and system administrators, external forces that seek to infiltrate these corporate networks, and so forth.

We assume that some insiders may be malicious and intentionally try to misuse the system and its mechanisms for creating trustworthiness. We also assume that external actors may seek to penetrate this system to then act internally in a similar manner as malicious insiders, and that insiders and external forces may collude in such activities. Here are some attack scenarios of interest:
\begin{itemize}
\item[S1] internal auditors may want to corrupt the state of the blockchain in order to cover up the traces of internal fraudulent activity

\item[S2] external forces may want to corrupt transaction hashes propagated in the internal network as a denial of service attack on the blockchain itself

\item[S3] control over a set of miners or nodes that propagate transaction hashes may be obtained so that certain transaction hashes have priority for mining

\item[S4] classical security attacks such as those on key management may be launched

\item[S5] a miner may amplify its computational power by sharing its private key and mining input with external computing resources.
\end{itemize}

Scenario S1 may be part of an insider attack in which skilled and sufficiently authorized insiders collude to commit fraud. The blockchain cannot ensure that only legitimate transactions have their hash recorded within it. But a \emph{security policy}~--~external to the blockchain~--~could specify that certain transactions are always recorded in the blockchain. For example, the hashes of security log entries (e.g.\ for the editing of security-relevant files) may also be added to the blockchain.

Scenario S2 is typical for a denial of service attack. The service that is being denied here is the blockchain, as the mechanism that creates sufficient trustworthiness into transactions as recorded within legacy system. These threats can be mitigated against, for example, legacy systems and the entire blockchain network may be placed behind a corporate firewall.

Scenario S3 is concerned with the management of transaction hashes that have not yet been recorded in the blockchain. The extent to which this is a security problem beyond that of
service availability will depend on the use context of the legacy accounts. For example, for the attack discussed in scenario S1 it might help internal attackers to have control over which transaction hashes would enter the blockchain first, with a preference of recording the fraudulent transaction that changed the recipient of payments of the originally recorded transaction.

In scenario S4, a cyber attack may get control of the part of the system that provides  authenticated information about the public keys of all used miners, for example, it might be able to learn a system admin key that allows for the modification of such information to change the private/public key pairs of miners to values of machines controlled by the attacker.

Scenario 5 can be mitigated against. For example, a firewall may block all communication that originates from miners if it does not propagate a Proof of Work on the internal network. Also, the private key of a miner may be stored in protected hardware so that it cannot be shared with other devices unless there is fraudulent activity in the supply chain or assembly of mining units.

Security would certainly be improved in these and other scenarios if all actions of blockchain management and of financial processes that modify accounts have their hashes recorded onto the blockchain, be these financial transactions, actions that create log entries, actions that give authority to perform financial transaction, and so forth. From this perspective, our blockchain could also facilitate a \emph{forensic} audit~--~not just one concerned with compliance and regulation.

\section{Mathematics for Centrally Governed Proof of Work}
\label{section:probabilisticmodel}
Our model assumes a cryptographic hash function \(h\colon \{0,1\}^p \to \{0,1\}^n\) where $p \geq n > 0$ such that $h$ has \emph{puzzle friendliness} \cite{narayanan16}. The
\emph{level of difficulty} $d$ is an integer satisfying $0 < d < n$: Proof of Work has to produce some $x$ where $h(x)$ has at least $d$ many leftmost $0$ bits.
We write $T > 0$ for the time to compute a sole hash $h(x)$ and to decide whether it has at least $d$ leftmost zeros. Since the range of $d$ will be relatively small, we make $T$ a device-dependent constant.

Our probabilistic modeling will treat $h$ in the \emph{Random Oracle Model} (ROM): function $h$ is chosen uniformly at random from all functions of type $\{0,1\}^p\to \{0,1\}^n$; that is to say, $h$ is a deterministic function such that any $x$ for which $h$ has not yet been queried will have the property that $h(x)$ is governed by a truly random probability distribution over $\{0,1\}^n$.

We may assume that $x$ consists of a block header which contains some random data field~--~a nonce $nonce$ of bitlength $r$, that this nonce is initialized, and that the nonce is then increased by $1$ each time the hash of $x$ does not obtain Proof of Work. In particular, this yields that \(\{0,1\}^p \cong \{0,1\}^{p-r}\times \{0,1\}^r\) where \(0 < r < p\): the input to $h$ will be of form $x = data\mid\mid nonce$ where $data$ and $nonce$ have $p-r$ and $r$ bits, respectively. Our use of ROM will rely on the assumption that mining, be it by a sole miner or in a mining race of more than one miner, will never revisit the same input again:

\begin{assumption}[Invariant]
\label{assumption:mininginvariant}
The mining of a block with one or more miners will use an input to $h$ at most once, be it within or across miners' input spaces.
\end{assumption}

This assumption and appeal to ROM give us that hash values are always uniformly distributed in the output space during a mining race. We now develop the probability space for mining with a sole miner, and then adapt this to the setting of more than one miner.

\subsection{Basic Probability Space for One Miner}  Our basic probability space has $data$ and $d$ as implicit parameters, and assumes the enumeration $0\dots 2^r-1$ of values of $nonce$ without loss of generality. The set of basic events $E$ of this probability space is

\begin{equation}
\label{equ:basicevents}
E = \{\fail^k\cdot \succeed\mid 0\leq k\leq 2^r-1\} \cup \{\failure\}
\end{equation}

\noindent where $\failure$ denotes the event that all $2^r$ nonce values failed to obtain Proof of Work for $data$ at level of difficulty $d$, and $\fail^k\cdot \succeed$ models the event in which the first $k$ such nonce values failed to obtain Proof of Work for $data$ at level $d$ but the $k+1$th value of $nonce$ did render such Proof of Work for $data$. We next want to define a discrete probability distribution \(prob\colon E\to [0,1]\) with mass $1$. Now, we have \(prob(\fail^0\cdot \succeed) = 2^{n-d} / 2^n = 2^{-d}\) since this is the fraction between the number of possible outputs that do Proof of Work at level of difficulty $d$ and the number of all possible outputs of $h$.

To define $prob(\fail^k\cdot \succeed)$ for the case when $k > 0$, we first need to understand the probability $\tilde{p}(\fail^k)$ of not obtaining Proof of Work for the first $k$ values of the nonce. For each value of $nonce$, the probability that $h(data\mid\mid nonce)$ does not have $d$ or more leading zeros is $1-prob(\fail^0\cdot\succeed) = 1-2^{-d}$. By ROM and Assumption~\ref{assumption:mininginvariant}, these probabilities are independent from each other for each of these $k$ different values of $nonce$ and so  $\tilde{p}(\fail^k) = (1-2^{-d})^k$ follows. Similarly, the probability that the $k+1$th hash attempt proves work is independent of whether or not any of the previous $k$ attempts did that~--~mining never tries that same nonce value again.
But then \(prob(\fail^k\cdot \succeed)$ is $(1-2^{-d})^k\cdot 2^{-d}\). We thus defined $prob(e)$ for events $e$ in $E\setminus \{\failure\}$ such that $0 < prob(e) < 1$. Also, the mass of all those probabilities is less than $1$:

\begin{eqnarray}
\sum_{e\in E\setminus \{\failure\}} \, prob(e) &=& \sum_{0\leq k\leq 2^r-1} (1-2^{-d})^k\cdot 2^{-d}\nonumber
\end{eqnarray}

\noindent but this equals $1 - (1-2^{-d})^{2^r} $ and is therefore in $(0,1)$ since $0 < d,r$. Thus, we obtain a probability distribution $prob$ by setting \[prob(\failure) = 1 - \sum_{0\leq k\leq 2^r-1} prob(\fail^k\cdot \succeed)\]

\noindent which equals
\(1 - (1 - (1-2^{-d})^{2^r}) = (1-2^{-d})^{2^r}\).

\subsection{Probability Space for $s > 1$ Miners} Consider having $s > 1$ many miners that run in parallel to find Proof of Work, engaging thus in a \emph{mining race}. We assume these miners run with the same configurations and hardware. In particular, the hash function $h$ and the values $n$, $p$, $d$, $r$, and $T$ will be the same for each of these miners. As already discussed, miners do not get rewarded:

\begin{assumption}[Miners]
\label{assumption:miners}
Miners are a resource controlled by the governing organization or consortium, and have identical hardware. In particular, miners are not rewarded nor have the need for incentive structures. \end{assumption}

But miners may be corrupted and misbehave, for example\ they may refuse to mine.
 To simplify our analysis, we assume miners begin the computation of hashes in approximate synchrony:

\begin{assumption}[Approximate Synchrony]
\label{assumption:synchronous}
Miners start a mining race at approximately the same time.
\end{assumption}

For many application domains, this is a realistic assumption as communication delays to miners would have a known upper bound that our models could additionally reflect if needed.

Next, we want to model the \emph{race} of getting a Proof of Work where each miner $j$ has some data $data_j$. To realize Assumption~\ref{assumption:mininginvariant}, it suffices that each miner $j$ have a nonce $nonce_j$ in a value space of size $\lfloor 2^r/s\rfloor$ such that these nonce spaces are mutually disjoint across miners.
To model this mining race between $s$ miners, we take the product \(\prod_{j=1}^s E^j\) of $s$ copies $E^j$ of our event space $E$ for mining with a sole miner, and quotient it via an equivalence relation $\equiv$ on that product \(\prod_{j=1}^s E^j\). The $s$-tuple $(\failure,\dots,\failure)$ models failure of this mining race, it is $\equiv$ equivalent only to itself.

All $s$-tuples $a = (a_j)_{1\leq j\leq s}$ other than tuple $(\failure,\dots,\failure)$ model that the mining race succeeded for at least one miner. For such an $s$-tuple $a$, the set  of natural numbers $k$ such that $\fail^k\cdot \succeed$ is a coordinate in $a$ is non-empty and therefore has a minimum $\min(a)$. Given two $s$-tuples $a = (a_j)_{1\leq j\leq s}$ and $b=(b_j)_{1\leq j\leq s}$ both different from $(\failure,\dots,\failure)$, we can then define $a$ and $b$ as $\equiv$ equivalent iff $\min(a)=\min(b)$. So two non-failing tuples are equivalent if they determine a first (and so final) Proof of Work at the same round of the race. This defines an equivalence relation $\equiv$ and adequately models a synchronized mining race between $s$ miners.

In the setting of $s > 1$ miners, the interpretation of events $\fail^k\cdot \succeed$ of $E$ in~(\ref{equ:basicevents}) is then the equivalence class of all those tuples $a$ for which $\min(a)$ is well defined and equals \(k\): all mining races that succeed first at round $k$. The meaning of $\failure$ is still overall failure of the mining race, the equivalence class containing only tuple $(\failure,\dots,\failure)$.

Next, we set
\[
\lambda = \lfloor 2^r/s \rfloor
\]

\noindent as the size of the nonce space for each of the $s$ miners, and define accordingly the set of basic events for $s$ miners as

\begin{equation}
\label{equ:Es}
E^{s} = \{\fail^k\cdot \succeed\mid 0\leq k\leq \lambda\} \cup \{\failure\}
\end{equation}

\noindent  In~(\ref{equ:Es}), expression $\fail^k\cdot \succeed$ denotes an element of the quotient $\bigl (\prod_{j=1}^s E^j \! \bigr )/ \!\!\equiv$, the equivalence class of tuple $(\fail^k\cdot \succeed,\failure,\failure,\dots,\failure)$. Also, $E^s$ restricts the set of non-failure events from $E$ in~(\ref{equ:basicevents}) to those with $k\leq \lambda$.

Next, we define a probability distribution $prob^s$ over $E^s$, consistent with the definition of  $prob$ over $E$ when $s$ equals $1$. To derive the probability $prob^{s}(\fail^k\cdot \succeed)$, recall $\tilde p(\fail^k) = (1-2^{-d})^k$ as the probability that a given miner does not obtain Proof of Work at level $d$ in the first $k$ rounds. By Assumption~\ref{assumption:mininginvariant}, these miners work independently and over disjoint input spaces. By ROM, the expression \(\bigl [ (1-2^{-d})^k \bigr ]^ s = (1-2^{-d})^{k\cdot s}\) therefore models the probability that none of the $s$ miners obtains Proof of Work in the first $k$ rounds. Appealing again to ROM and Assumption~\ref{assumption:mininginvariant}, the behavior at round $k+1$ is independent of that of the first $k$ rounds. Therefore, we need to multiply the above probability with the one for which at least one of the $s$ miners will obtain a Proof of Work in a single round. The latter probability is the complementary one of the probability that none of the $s$ miners will get a Proof of Work in a sole round, which is $(1-2^{-d})^s$ due to the ROM independence. Therefore, we get
\begin{equation}
prob^{s}(\fail^k\cdot \succeed) = (1-2^{-d})^{k\cdot s}\cdot [1-(1-2^{-d})^s]
\end{equation}

This defines a probability distribution with a non-zero probability of $\failure$. Firstly, \(\sum_{k=0}^{\lambda} (1-2^{-d})^{k\cdot s}\cdot [1-(1-2^{-d})^s]\) is in $(0,1)$: that sum equals \[[1-(1-2^{-d})^s]\cdot\frac{1-[(1-2^{-d})^s]^{\lambda+1}}{1-(1-2^{-d})^s} =1-(1-2^{-d})^{s\cdot(\lambda+1)}\]

\noindent And since $0 < d,s$, the real $1-2^{-d}$ is in the open interval $(0,1)$, and the same
is true of any integral power thereof. Secondly, $prob^s$ becomes a probability distribution with the non-zero probability $prob^s(\failure)$ being $1-prob^e(E^s\setminus \{\failure\})$, that is
\begin{equation}
prob^s(\failure) = (1-2^{-d})^{s\cdot(\lambda+1)}
\end{equation}

\noindent This failure probability is almost identical to that for $s=1$, an artefact of our parameter representation: for example, if each miner has $64$ bits of nonce space, then our model would have $r = 64\cdot s$, so failure probabilities do decrease as $s$ increases.

\section{Mathematical Optimization in Mining Design Space}
\label{section:optimization}

\subsection{Generality of Approach}
We want to optimize the use of $s > 1$ miners using a level of difficulty $d$, and a bit size $r$ of the global nonce space with respect to an objective function. The latter may be a cost function, if containing cost is the paramount objective or if a first cost estimate is sought that can then be transformed into a constraint to optimize for a security objective, as seen further below.

Higher values of $d$ add more security: it takes more effort to mine a block and so more effort to manipulate the mining process and used consensus mechanism. But lower values of $d$ may be needed, for example, in high-frequency trading where performance can become a real issue. We want to understand such trade-offs.

Moreover, we want to explore how the corruption of a number of miners or inherent uncertainty in the number of deployed miners or in the level of difficulty across the lifetime of a system may influence the above tradeoffs. We will use tools from robust optimization~\cite{bental09} and functional programming to analyze such issues.

\subsection{Optimizing Cost and Security}
The flexibility of our approach includes the choice of objective function for optimization. Let us first consider an objective function
\begin{equation}
\label{equ:costfunction}
\mbox{Cost}(s,r,d) = \mbox{TVC}\cdot E^s(noR)\cdot s + \mbox{TFC}\cdot
s
\end{equation}

\noindent that models cost as a function of the number of miners $s$, the bit size of the nonce $r$~--~implicit in random variable $E^s(noR)$, and the level of difficulty $d$; where we want to \emph{minimize} cost.

The real variable $\mbox{TVC}$ models the \emph{variable} cost of computing \emph{one} hash for \emph{one} miner, reflecting the device-dependent speed of hashes and the price of energy. The real variable $\mbox{TFC}$ models the \emph{fixed} costs of \emph{having one miner}; this can be seen as modeling procurement and depreciations. Variables $s$, $r$, and $d$ are integral, making this a \emph{mixed integer} optimization problem \cite{DBLP:books/daglib/0023873}. The expression $E^s(noR)$ denotes the \emph{expected number of rounds} needed to mine a block in a mining race that uses $s$ miners, level of difficulty $d$, and nonce bitsize $r$. The derivation of this expression below shows that it is non-linear, making this a MINLP optimization problem \cite{gamsworld01,DBLP:books/daglib/0023873}. We chose not to include in $\mbox{TVC}$ a constant that reflects how many blocks may be mined within a system horizon of interest but this can easily be done within our proposed approach, for example when there is a constraint on the carbon footprint of a system during its lifetime.

We may of course use other objective functions. One of these is simply the expression $d$, which we would seek to \emph{maximize}, the intuition being that higher values of $d$ give us more trust into the veracity of a mined block and the blockchains generated in the system.
Figure~\ref{fig:firstopt} shows an example of a set of constraints and optimizations of security and cost for this.
\begin{figure}
\begin{eqnarray*}
0 &<& s_l \leq s\leq s_u\qquad 0 < d_l \leq d\leq d_u\qquad 0 < r_l \leq r\leq r_u\qquad \epsilon \geq prob^s(\failure) \\
\tau_u &\geq& T\cdot E^s(noR) \geq \tau_l \qquad \qquad \ \ \ \delta_2\geq prob^s(disputes\ within\ \mu) \\
\delta &\geq& prob^s(PoWTime > th) \qquad \delta_1 \geq prob^s(PoWTime < th') 
\end{eqnarray*}
\caption{Constraint set $\mathcal C$ for two optimization problems: (a) \emph{minimize} $\mbox{Cost}(s,r,d)$ as in~(\ref{equ:costfunction}) subject to constraints in $\mathcal C$; and (b)  \emph{maximize} $d$ subject to 
  ${\mathcal C}\cup \{\mbox{Cost}(s,r,d)\leq budget\}$ for cost bound $budget$. This is parameterized by constants $0 \leq \delta, \delta_1, \delta_2, \epsilon, th, th', \tau_l, \mbox{TVC}, \mbox{TFC}$ and $0 < T, s_l, r_l, d_l$. Variables or constants $s_l, s_u, s, d_l, d_u, d, r_l, r_u, r$ are integral\label{fig:firstopt}}
\end{figure}

Integer constants $s_l$ and $s_u$ provide bounds for variable $s$, and similar integer bounds are used to constrain integer variables $r$ and $d$. The constraint for $\epsilon$ uses it as upper bound for the probability of a mining race failing to mine a block. The next two inequalities stipulate that the expected time for mining a block is within a given time interval, specified by real constants $\tau_l$ and $\tau_u$.

The real constant $\delta_2$ is an upper bound for $prob^s(disputes\ within\ \mu)$, the probability that more than one miner finds PoW within $\mu$ seconds in the same, synchronous, mining race.
The constraint for real constant $\delta$ says that the probability $prob^s(PoWTime > th)$ of the \emph{actual} time for mining a block being above a real constant $th$ is bounded above by $\delta$. This
constraint is of independent interest: knowing that the expected time to mine a block is within specified bounds may not suffice in systems that need to assure that blocks are \emph{almost always} (with probability at least $1-\delta$) mined within a specified time limit.
Some systems may also need assurance that blocks are almost always mined in time \emph{exceeding} a specified time limit $th'$. We write $prob^s(PoWTime < th')$ to denote that probability, and add a dual constraint specifying that the actual time for mining a block has a sufficiently small probability $\leq \delta_1$ of being faster than some given threshold $th'$.

\subsection{Constraints as Analytical Expressions}
We derive analytical expressions for random variables
occurring in
Figure~\ref{fig:firstopt}. Beginning with
$E^s(noR)$,
we have
\(E^s(noR) = \sum_{0\leq k\leq \lambda} prob^s(\fail^k\cdot
\succeed)\cdot (k+1)\) which we know to be equal to
\(\sum_{0\leq k\leq \lambda}  (1-2^{-d})^{k\cdot s}\cdot [1-(1-2^{-d})^s]\cdot (k+1)\).
We may rewrite the latter expression so that summations
are eliminated and reduced to exponentiations: concretely, we rewrite
\(\sum_{0\leq k\leq \lambda} prob(\fail^k\cdot \succeed)\cdot (k+1)\) to
$\lambda+1$ summations, each one starting at a value between $0$ and
$\lambda$, where we exploit \(\sum_{k=a}^b x^k= \frac{x^a - x^{b+1}}{1-x}\).
This renders
\begin{equation}
\label{equ:noranalytical}
E^s(noR) = \frac{1-y^{\lambda+1} - (\lambda+1)\cdot
  (1-y)\cdot y^{\lambda+1}}{1-y}
\end{equation}

\noindent where we use the abbreviation
\begin{equation}
\label{equ:y}
y = (1-2^{-d})^s
\end{equation}

\noindent The expected time needed to get a proof of
work for input $data$ is then given by
\begin{equation}
E^s(poW) = T\cdot E^s(noR)
\end{equation}

We derive an analytical expression for the above probability $prob^s(PoWTime > th)$ next.
 Note that
 $(th/T)-1< k$ models that the actual time
taken for $k+1$ hash rounds is larger than $th$. Therefore, we
capture $prob^s(PoWTime > th)$ as
\begin{eqnarray}
\label{equ:moretimes}
\sum_{\lceil (th/T) - 1\rceil < k\leq \lambda} prob^s(\fail^k\cdot \succeed) &=& {} \\
\sum_{\lceil (th/T) - 1 \rceil < k\leq \lambda}  (1-2^{-d})^{k\cdot s}\cdot [1-(1-2^{-d})^s] &=& {}\nonumber\\
(1-2^{-d})^{s\cdot (\lceil (th/T) -1\rceil + 1)} - (1- 2^{-d})^{s\cdot (\lambda+1)} &=& {}\nonumber\\
y^{\lceil (th/T) -1\rceil + 1} - y^{\lambda+1} &{}&\nonumber
\end{eqnarray}

\noindent assuming that $\lceil (th/T) - 1\rceil < \lambda$, the
latter therefore becoming a constraint that we need to add to our
optimization problem. One may be tempted to choose the value of
$\delta$ based on the Markov inequality, which gives us
\[
prob^s(PoWTime \geq th)\leq T\cdot E^s(noR)/th
\]

\noindent But we should keep in mind that
upper bound $T\cdot E^s(noR)/th$ depends on the parameters $s$, $r$, and $d$; for example, the analytical expression for $E^s(noR)$ in~(\ref{equ:noranalytical}) is dependent on $\lambda$ and so dependent on $r$ as well. The representation in~(\ref{equ:moretimes}) also maintains that expression %
\[
y^{\lceil (th/T) -1\rceil + 1} - y^{\lambda+1}
\]

\noindent is in $[0,1]$, i.e.\ a proper probability. Since $y=(1-2^{-d})^s$ is in $(0,1)$, this is already guaranteed if
$\lceil (th/T) -1\rceil + 1\leq \lambda+1$, i.e.\ if $\lceil (th/T) -1\rceil \leq \lambda$. But we already added that constraint to our model.
Similarly to how we proceded for $prob^s(PoWTime > th)$, we get
\begin{equation}
\label{equ:lesstimeprobs}
prob^s(PowTime < th') =  1 - (1-2^{-d})^{s\cdot (\lfloor (th'/T) -1\rfloor+1)} = 1- y^{\lfloor (th'/T) -1\rfloor+1}
\end{equation}

\noindent which needs $0 < \lfloor (th'/T) - 1\rfloor$ as
additional constraint.

To derive an analytical expression for  $prob^s(disputes\ within\ \mu)$, each miner can perform  $\lfloor \mu/T \rfloor$ hashes within $\mu$ seconds. Let us set
\begin{equation}
w =  (1-2^{-d})^{\lfloor \mu/T \rfloor + 1}
\end{equation}

\noindent The probability that a given miner finds PoW within $\mu$ seconds is
\begin{equation}
\sum_{k=0}^{\lfloor \mu/T \rfloor} (1-2^{-d})^k\cdot 2^{-d} = 2^{-d}\cdot \frac{1-(1-2^{-d})^{\lfloor \mu/T \rfloor + 1}}{1-(1-2^{-d})}
= 1-w
\end{equation}

\noindent Therefore, the probability that no miner finds PoW within $\mu$ seconds is
\begin{equation}
prob^s(0\ PoW\ within\ \mu) = (1 -  (1- w))^s = w^s
\end{equation}

\noindent The probability that exactly one miner finds PoW within $\mu$ seconds is
\begin{equation}
prob^s(1\ PoW\ within\ \mu) = s\cdot w^{s-1} \cdot (1-w)
\end{equation}

\noindent Thus,  the probability that more than one miner finds PoW within $\mu$ seconds is
\begin{eqnarray}
prob^s(disputes\ within\ \mu) &=& 1 - prob^s(0\ PoW\ within\ \mu) - prob^s(1\ PoW\ within\ \mu)\nonumber\\
{} &=& 1 - w^s - s\cdot w^{s-1}\cdot (1-w)\nonumber\\
{} &=& 1 - w^s -s\cdot w^{s-1} + s\cdot w^{s-1}\cdot w\nonumber\\
{} &=& 1 + (s-1)\cdot w^s - s\cdot w^{s-1}
\end{eqnarray}
Figure~\ref{fig:firstoptexplicit} shows the set of constraints
$\mathcal C$ from Figure~\ref{fig:firstopt} with analytical
expressions and their additional constraints, we add constraint $0\leq \lfloor \mu/T \rfloor$ to get consistency for the analytical representation of $prob^s(disputes\ within\ \mu)$.
\begin{figure*}
\begin{eqnarray}
s_l &\leq& s\leq s_u\qquad d_l \leq d\leq d_u \qquad r_l \leq r\leq r_u\qquad \lambda = \lfloor 2^r/s \rfloor\nonumber\\
y &=& (1-2^{-d})^s \qquad w =  (1-2^{-d})^{\lfloor \mu/T \rfloor + 1} \qquad 0\leq \lfloor \mu/T \rfloor \nonumber\\
\epsilon &\geq& y^{\lambda+1}  \qquad \lceil (th/T) - 1\rceil < \lambda \qquad 0 < \lfloor (th'/T) - 1\rfloor \nonumber\\
 E^s(noR) &=& \frac{1- y^{\lambda+1} - (\lambda+1)\cdot
  (1-y)\cdot y^{\lambda+1}}{1-y} \nonumber\\
\tau_u &\geq& T\cdot E^s(noR) \geq \tau_l \qquad \delta_1 \geq 1 - y^{\lfloor (th'/T) -1\rfloor+1} \nonumber\\
\delta &\geq& y^{\lceil (th/T) -1\rceil + 1} - y^{\lambda+1} \label{equ:deltaprime} \\
\delta_2 &\geq& 1 + (s-1)\cdot w^s - s\cdot w^{s-1}\nonumber
\end{eqnarray}
\caption{Arithmetic version of set of constraints $\mathcal C$
from Figure~\ref{fig:firstopt}, with additional soundness constraints for this representation.
 Feasibility of $(s,r,d)$ and $r_u\geq r' > r$ won't generally imply feasibility of $(s,r',d)$ due to the constraint in~(\ref{equ:deltaprime}) \label{fig:firstoptexplicit}}
\end{figure*}

\subsection{Robust Design Security}
\label{section:security}
Our model above captures design requirements or design decisions as a set of constraints, to optimize or trade off measures of interest subject to such constraints.
We can extend this model to also \emph{manage uncertainty} via robust optimization~\cite{bental09}. Such uncertainty may arise during the lifetime of a system through the possibility of having corrupted miners, needed flexibility in adjusting the level of difficulty, and so forth.
For example, corrupted miners may refuse to mine, deny their service by returning
invalid block headers, pool their mining power to get more mining influence or
they may simply break down.

Consider $1\leq l < s$ corrupted miners. We can model their \emph{pool power} by appeal to ROM and the fact that the mining race is roughly synchronized: the probability that these $l$ miners win $c > 0$ many subsequent mining races is then seen to be $(l/s)^c$. We can therefore bound this with a constant $\delta_3$, or equivalently we can add to the set of constraints $\mathcal C$ from Figure~\ref{fig:firstoptexplicit} the constraint $l^c\leq \delta_3\cdot s^c$.

We model uncertainty in the number of miners available by an integer
constant $u_s$ as follows: if $s$ miners are deployed, then we assume
that at least $s-u_s$ and at most $s$ many miners participate reliably in the
mining of legitimate blocks: they will not mine blocks that won't verify and only submit mined blocks that do verify to the network. Therefore, $u_s$ allows us to model
aspects such as denial of service attacks or a combination of such
attacks with classical faults: for example, $u_s=3$ subsumes the scenario in
which one miner fails and two miners mine invalid blocks.

Furthermore, an integer constant $u_d$ models the uncertainty we have
in the deployed level of difficulty $d$: the intuition is that our
analysis should give us results that are robust in that they hedge
against the fact that any of the values $d'$ satisfying
$\size{d-d'}\leq u_d$ may be the actually running level of
difficulty. This enables us to understand a design if we are unsure
about which level of difficulty will be deployed or
if we want some flexibility in dynamically adjusting the value of $d$ in the running system.

The corresponding robust optimization problem for cost minimization is seen in
Figure~\ref{fig:robustcost}. It adds to the constraints we already
consider the requirements on constants $l$, $c$, and $\delta_3$ as
well as the constraint $l^c\leq \delta_3\cdot s^c$. This problem chooses values for $c$, $l$, $u_d$, and $u_s$ for sake of concreteness. The robustness of
analysis is achieved by a change of the objective function from
$\mbox{Cost}(s,r,d)$ to
\begin{equation}
\label{equ:robustcost}
\mbox{Cost}^{u_s}_{u_d}(s,r,d) = \max_{s-u_s\leq s'\leq s, \,\size{d-d'}\leq u_d} \mbox{Cost}(s',r,d')
\end{equation}

\noindent The latter computes a worst-case cost for triple $(s,r,d)$
where $s$ and $d$ may vary independently subject to the strict
uncertainties $u_s$ and $u_d$, respectively. We call a triple $(s,r,d)$ \emph{feasible} if it satisfies all constraints of its optimization problem.
Costs such as the one in~(\ref{equ:robustcost}) for
a triple $(s,r,d)$ are only considered for optimization if all triples
$(s',r,d')$ used in the robust cost computation in~(\ref{equ:robustcost}) are
feasible~--~realized with predicate $f\!easible^{u_s}_{u_d}$: robust optimization guarantees \cite{bental09} that the feasibility of
solutions is invariant under the specified uncertainty (here $u_s$ and $u_d$).
\begin{figure*}[h]
\begin{eqnarray*}
{}&{}& \min \{ \mbox{Cost}^{u_s}_{u_d}(s,r,d) \mid
f\!easible^{u_s}_{u_d}(s,r,d) \} \\
{} &{}& \mbox{subject to the set of constraints $\mathcal C$ from
  Figure~\ref{fig:firstoptexplicit} together with} \\ 
{} &{}& 4 = l < s\qquad\qquad c = 6\qquad\qquad 0.001 = \delta_3\\
{} &{}& l^c \leq s^c\cdot \delta_3 \qquad\qquad u_s = 5\qquad\qquad
u_d = 3
\end{eqnarray*}
\caption{Robust optimization that minimizes cost for the set of
  constraint from
Figure~\ref{fig:firstoptexplicit},
  where up to $u_s=5$ miners may be either non-functioning, refusing to
  mine or mining invalid blocks; where the level of difficulty may
  vary by up to $+3$ or $-3$; and where we want that the
  probability of any mining \emph{pool} of size $l=4$ winning $c=6$
  consecutive mining races is sufficiently small (here $\delta_3 = 0.001$). Predicate
  $f\!easible^{u_s}_{u_d}(s,r,d)$ characterizes \emph{robustly feasible} triples and is true iff all triples $(s',r,d')$
  with $s-u_s\leq s'\leq s$ and $\size{d-d'}\leq u_d$ are feasible\label{fig:robustcost}}
\end{figure*}

\section{Experiments and Validation}
\label{section:experiments}
We submitted simple instances of the optimization problem in Figure~\ref{fig:robustcost} to state of the art MINLP solvers. All these solvers reported, erroneously, in their preprocessing stage that the problem is infeasible. These solvers were not designed to deal with problems that combine such small numbers and large powers, and rely on standard floating point implementations. Therefore, we wrote a bespoke solver in Haskell that exploits the fact that we have only few integral variables within limited ranges so that we can explore their combinatorial space completely to determine feasibility.
\begin{figure}[ht!]\label{fig:algorithm}
\setlength{\interspacetitleruled}{-.4pt}%
\begin{algorithm}[H]
\SetKwData{Left}{left}\SetKwData{This}{this}\SetKwData{Up}{up}
\SetKwFunction{Union}{Union}\SetKwFunction{FindCompress}{FindCompress}
\SetKwInOut{Input}{input}\SetKwInOut{Invariant}{invariant}
\Indm
\Input{$p$, $\alpha$, and values for all constants in Figure~\ref{fig:robustcost}}
\Indp
\Indm
\Invariant{$list$ lists tuples $(s,r,d,cost)$ in descending order for $d$}
\Indp
\Begin{
$de\!f\!ine\ all\ constants\ f\!or\ constraints\ in\ Figure~\ref{fig:robustcost}$\;
$list = [(s,r,d,cost) \mid cost = Cost(s,r,d), f\! easibleFloat(s,r,d)\ is\ true]$\;
$list = [(s,r,d,cost)\in list \mid f\! easibleFloat^{u_s}_{u_d}(s,r,d)\ is\ true ]$\;
\While{$(\exists (s,r,d,cost)\neq (s',r',d,cost')\in list)$}{
\lCase{$(cost' < cost)\lor (r' < r)$}{$remove\ (s,r,d,cost)\ f\! r\! om\ list$)}
\lCase{$(cost < cost')\lor (r < r')$}{$remove\ (s',r',d,cost')\ f\! r\! om\ list$)}
}
$c_m = \min \{ c\mid \exists (s,r,d,cost)\in list \}$\;
\While{$(\exists (s,r,d,cost)\in list\colon cost > \alpha\cdot c_m)$}{
$remove\ (s,r,d,cost)\ f\! r\! om\ list$\;
}
$results = list\ o\! f\ f\! ir\! st\ p\ tuples\ f\! r\! om\ list$\;
$results = [ (s,r,d,cost)\in results \mid f\! easibleBigFloat^{u_s}_{u_d}(s,r,d)\ is\ true]$\;
\Return{results\;}
}
\end{algorithm}
\caption{Algorithm, written in imperative style of list processing, for reporting the best $p$ robustly feasible tuples $(d,r,s,cost)$ such that $d$ is maximal subject to the cost $cost = Cost(s,r,d)$ satisfying $cost\leq \alpha\cdot c_m$ where $c_m$ is the minimal cost for all robustly feasible tuples $(s,r,d)$ and $\alpha \geq 1$ is a tolerance factor for increasing cost beyond $c_m$. Predicate $f\! easibleFloat(s,r,d)$ is true iff all constraints in Figure~\ref{fig:firstoptexplicit} are true for this choice of $s$, $r$, and $d$ under normal precision floats. Predicates $f\! easibleBigFloat$ and $f\! easibleBigFloat^{u_s}_{u_d}$ are true iff their mathematical definition is true under arbitrary-precision floating points (applying ${\tt Data.BigFloat}$ version 2.13.2).}
\end{figure}

\subsection{Experimental Setup}
We solve the robust optimization problem for the analytical expressions we derived above with the algorithm depicted in Figure~\ref{fig:algorithm}. This algorithm has as input the set of constraints, a parameter $p$ and a parameter $\alpha$. It will output at most $p$ robustly feasible tuples $(s,r,d,cost)$ from a list of all robustly feasible such tuples as follows: it will identify the maximal values of $d$ for which such tuples are robustly feasible, and it will report exactly one such tuple for each value of $d$ where $r$ is minimal, and $cost$ is minimal whilst also bounded above by $\alpha\cdot c_m$ where $c_m$ is the globally minimal cost. This also determines the values of $s$ in these tuples and so the algorithm terminates.

Now, having defined the required analytical expressions and the algorithm to report the best $p$ robustly feasible tuples in Figure~\ref{fig:algorithm}, we also want to validate these expressions and the algorithm experimentally. Our setup for this is based on pure Haskell code, as functional~--~and in particular~--~Haskell programs offer the advantages of being modular in the dimension of functionality, being strongly typed as well as supporting an easy deconstruction of data structures, particularly lists~\cite{Bird2015}. Furthermore, the arbitrary-precision verification is handled by the external ${\tt Data.BigFloat}$ package, which is also written in Haskell. Further verification and validation of the received results is pursued by unit testing using an arbitrary precision calculator. Moreover, our experiments ran on a machine with the following specifications: Intel(R) Xeon(R) CPU E5-4650 with 64 cores and 2.70GHz and 500 GB overall RAM. Our machines required between 322.12 and 261.425 seconds to compute the respective optimizations. The entire experiment took 10457.58 seconds.
\begin{table}
\[
\begin{array}{|l|l|l|l|}
\hline
s_l       = 4 & s_u       = 80 & r_l       = 24 & r_u       = 64 \\
d_l       = 4 & d_u       = 64 & \mbox{TVC} = 2\cdot 10^{-12} & \mbox{TFC} = 3000 \\
\alpha     = 1.5 &   T      = 0.002\cdot 10^{-9}  & th        = 300 & th' = 300 \\
\delta     = 10^{-9} & \delta_1    = 1 & \delta_3   = 0.001 & \delta_2 = 0.001 \\
\tau_l     = 0 & \mu = 1/10000 & \epsilon   = 2^{-64} &  k         = 5\\
u_d       = 3 & u_s       = 5 & c         = 6 & l         = 4 \\
\hline
\end{array}
\]
\caption{Constants for our experiments. This does not specify the values of $\tau_u$ which will vary in experiments. Some experiments will also vary the values of $\delta$, $\delta_2$ or $\delta_3$\label{table:const}}
\end{table}

We instantiate the model in Figure~\ref{fig:robustcost} with the constants shown in Table~\ref{table:const}.  We choose $T$ to be $1/(50\cdot 10^9) = 0.02\cdot 10^{-9}$ for a
mining ASIC from early 2016 with an estimated cost of 2700 USD at that
time, so a fixed cost of $\mbox{TFC} = 3000$ USD seems reasonable. Let us now explain
the value $2\cdot 10^{-12}$, which models the energy cost of a sole
hash (we can ignore other costs on that time scale). A conservative
estimate for the power consumption of an ASIC is $10$ watts per
Gigahashes per second, i.e.\ $10$ watts per $Gh/s$. We estimate the
cost of one kilowatt hour $kWh$ to be about $10$ cents. A $kWh$ is
$3600s$ times $kW$ and one $kW$ is $1000$ watts. So $10$ watts per
$Gh/s$ equals $10\cdot 3600$ watts, which amounts to $36kWh$. So the
cost for this is $36\cdot 10$ cents per hour, i.e. $360$ cents per
hour. But then this costs $360/3600 = 0.1$ cents per second. The price
for a sole hash is therefore $0.1$ divided by $50\cdot 10^{9}$, which
equals $\mbox{TVC} = 2\cdot 10^{-12}$.

We insist on having at least $4$ miners and cap this at $80$ miners. The \emph{shared} nonce space for miners is assumed to be between $24$ and $64$ bits. The level of difficulty is constrained to be between $4$ and $64$. We list optimal tuples that are within a factor of $\alpha=1.5$ of the optimal cost. We make the value $th'$ irrelevant by setting $\delta_1 = 1$ which makes the constraint for $th'$ vacuously true. The probability for mining failure is not allowed to exceed $\epsilon=2^{-64}$. Setting $\tau_l=0$ means that we don't insist on the average mining time to be above any particular positive time. The probability that mining a block takes more than $th=300$ seconds is bounded by $10^{-9}$. And the probability that more than one miner finds PoW within $\mu = 1/10000$ seconds is bounded by $0.001$, which we also take as an upper bound for winning $6$ consecutive mining races. The algorithm reports the top $k=5$ optimal tuples~--~and reports fewer if there are no $5$ feasible tuples. The remaining constants for robustness are as given in Figure~\ref{fig:robustcost}.

Let us now specify some values of $\tau_u$ of interest.
As reported in \cite{DBLP:journals/corr/DanezisM15}, Bitcoin is believed to handle
up to $7$ transactions per second (although this can be improved \cite{Gervais16}), Paypal at least $100$ transactions
per second (which we take as an average here), and Visa anywhere
between $2000$ and $7000$ transactions per second on average.
By
\emph{transactions per second} we mean that blocks are mined within a
period of time consistent with this. Of course, this depends on
how many transactions are included in a block. For sake of
concreteness and illustration, we take an average number of transactions in a Bitcoin block, as reported for the beginning of April 2016, that is $1454$ transactions.

For a Bitcoin style rate, but in our \emph{governed} setting, this means that a block is mined in about $1454/7 \sim 207.71$ seconds. Since $T\cdot E^s(noR)$ is the expected
(average) time to mine a block, we can model that we have $7$
transactions per second on average by setting $\tau_u^{Bitcoin}$ to be $1454/7$. Similarly, we may compute $\tau_u^{PayPal}$ and $\tau_u^{Visa}$ based on respective $100$ and $7000$ transactions per second:
\begin{equation}
\label{equ:taus}
\tau_u^{Bitcoin} = 1454/7\qquad \tau_u^{PayPal} = 1454/100\qquad \tau_u^{Visa} = 1454/7000
\end{equation}

\subsection{Experimental Results}
We now discuss the results of our experiments. Each experiment is conducted in three different configurations:
\begin{itemize}
\item[C1] constants in as Table~\ref{table:const}, i.e.\ $\delta=10^{-9}$, $\delta_2 = \delta_3 = 0.001$
\item[C2] smaller $\delta$, that is $\delta=2^{-64}$, $\delta_2 = \delta_3 = 0.001$ 
\item[C3] smaller $\delta$ and $\delta_3$, that is $\delta=2^{-64}$, $\delta_2 = 0.001$, and $\delta_3 = 0.0001$. 
\end{itemize}

\paragraph{Transactions per second as in Bitcoin, PayPal, and Visa.} We show in Table~\ref{table:firstresults} output for the top $5$ optimal robustly feasible tuples for the various values of $\tau_u$ in~(\ref{equ:taus}) for configuration C1. We see that all three transaction rates can be realized with $18$ miners and a $48$-bit shared nonce space in our governed setting, and this gives each miner a nonce space of about $43$ bits.
 The achievable level of difficulty (within the uncertainty in $u_s$ and $u_d$) ranges from $37$ to $41$ for both the Bitcoin style rate and the PayPal style rate. For the Visa style rate, the feasible levels of difficulty are $34$ and $35$. For the optimal tuples reported in Table~\ref{table:firstresults}, the value of $r$ remains feasible whenever $48 \leq r\leq 64$. Note that these results also imply that, for all three rate styles, feasibility requires at least $18$ miners.
\begin{table}
{\small
\[
\begin{array}{|l|ll|ll|}
\hline
\tau_u^{Bitcoin} \ (s,r,d,cost) & \ \ &\tau_u^{PayPal} \ (s,r,d,cost) & \ \ &\tau_u^{Visa} \ (s,r,d,cost)\\
\hline
 \ (18,48,41,54004.4) & \ \ & (18,48,41,54004.4)  & \ \ & (18,48,35,54000.07)  \\
 \  (18,48,40,54002.2) & \ \ & (18,48,40,54002.2) & \ \ & (18,48,34,54000.035)  \\
 \ (18,48,39,54001.1) & \ \ &  (18,48,39,54001.1)  & \ \ & \\
\  (18,48,38,54000.55) & \ \ & (18,48,38,54000.55) & \ \ & \\
\   (18,48,37,54000.27) & \ \ & (18,48,37,54000.27)  & \ \ & \\
\hline
\end{array}
\]
} 
\caption{Output for top $5$ optimal tuples for our robust optimization problem run in configuration C1  and with values $\tau_u$ as listed in~(\ref{equ:taus}): $5$ optimal tuples are found for $\tau_u^{Bitcoin}$ and $\tau_u^{PayPal}$, i.e.\ at least $5$ values of $d$ are feasible. The problem has two feasible levels of difficulty for $\tau_u^{Visa}$. Costs are rounded up for three decimal places \label{table:firstresults}}
\end{table}

Let us run this experiment in configuration C2.
This models that the probability of mining to take more than $300$ seconds is very small.
We now only report the changes to the results shown in Table~\ref{table:firstresults} for the top rated, optimal tuple. For $\tau_u^{Bitcoin}$, the level of difficulty drops from $41$ to $40$ but there are still $18$ miners and a shared nonce space of $48$ bits. This tuple $(s,r,d) = (18,48,40)$ is also optimal for $\tau_u^{PayPal}$ now, whereas the optimal tuple $(s,r,d) = (18,48,35)$ for $\tau_u^{Visa}$ from configuration C1 remains to be optimal for C2. 

Next, we run this experiment for configuration C3, also decreasing the probability that corrupt miners can win $6$ consecutive mining races. For $\tau_u^{Bitcoin}$ and for $\tau_u^{PayPal}$, the top $5$ optimal tuples are $(s,r,d) = (24,49,d)$ where $36\leq d\leq 40$. In particular, this requires at least one more bit for the nonce space and at least $6$ more miners. For $\tau_u^{Visa}$, only tuples $(24,49,35)$ and $(24,49,34)$ are reported, so this also requires at least $24$ miners and a $49$-bit nonce space, where $35$ and $34$ are the feasible levels of difficulty.

We may explore the \emph{feasibility boundary} for $\tau_u$ for configuration C2.
The robust optimization problem is infeasible for $\tau_u=0.06871$ but becomes feasible when $\tau_u$ equals $0.06872$. In that case, the only feasible tuples are $(s,r,d) = (18,r,34,54000.03)$ where $48\leq r\leq 64$.

\paragraph{Larger transaction rates per second.} Next, we want to vary the average number of transactions $ant$ in a block from $ant=1454$ to larger values. This is sensible for our use case as transactions only record a hash, which may be $8$ bytes each. These results are seen in Table~\ref{table:secondresults} for $50000$ transactions on average in a block, running in the configuration C1. Let us discuss the impact of changing the $ant$ in a block from $1454$ to $50000$. This has no impact when $7$ or $100$ transactions per second are desired. For $7000$ transactions per second, this robust optimization problem still has the same $s$ and $r$ values in optimal tuples but the level of difficulty (which was $35$ or $34$) can now be between $36$ and $40$.
This quantifies the security and availability benefits from packing more transactions into a block for mining throughput.
\begin{table}
{\small
\[
\begin{array}{|l|ll|ll|}
\hline
Bitcoin\equiv 7 \ (s,r,d,cost) & \ \ &PayPal\equiv 100 \ (s,r,d,cost) & \ \ &Visa\equiv 7000 \ (s,r,d,cost)\\
\hline
\ (18,48,41,54004.4) & \ \ & (18,48,41,54004.4)  & \ \ &  (18,48,40,54002.2)  \\
 \  (18,48,40,54002.2) & \ \ & (18,48,40,54002.2) & \ \ &  (18,48,39,54001.1) \\
 \ (18,48,39,54001.1) & \ \ &  (18,48,39,54001.1)  & \ \ & (18,48,38,54000.55) \\
\  (18,48,38,54000.55) & \ \ & (18,48,38,54000.55) & \ \ & (18,48,37,54000.27)  \\
\   (18,48,37,54000.27) & \ \ & (18,48,37,54000.27)  & \ \ & (18,48,36,54000.138) \\
\hline
\end{array}
\]
} 
\caption{Output for top $5$ optimal tuples for our robust optimization problem running in configuration C1 and with values $\tau_u$ given as $50000/7$, $50000/100$, and $50000/7000$ (respectively). Results for the first two columns are identical with those in the first two columns of Table~\ref{table:firstresults}. The first $4$ optimal tuples for $\tau_u=  50000/7000$ equal that last $4$ of the $5$ optimal tuples for $50000/7$. Costs are rounded up for three decimal places \label{table:secondresults}}
\end{table}

Let us now see how these results change when we run the experiment in configuration C2. Now, all three rate styles report the same optimal $5$ tuples which are equal to the tuples listed in the rightmost column in Table~\ref{table:secondresults}: $(s,r,d) = (18,48,d)$ where $36\leq d\leq 40$.
The results for configuration C3 are also identical for all three rate styles, they equal $(s,r,d) = (24,49,d)$ where $36\leq d\leq 40$. So this requires one more bit in the nonce space and at least $6$ more miners.

\paragraph{Feasibility boundary for transaction rates per second.} We repeat the last experiment by varying the $ant$ from $50000$ to half a million, in increments of $50000$. We summarize these results as follows:
\begin{itemize}
\item \emph{Configuration C1:} For all three rate styles and all transaction values in increments of $50000$ up to $500000$, the optimal tuples are the same: $(s,r,d) = (18,48,d)$ where $37\leq d\leq 41$.

\item \emph{Configuration C2:} For all three rate styles and all transaction values in increments of $50000$ from $100000$ up to $500000$, the optimal tuples are the same: $(s,r,d) = (18,48,d)$ where $36\leq d\leq 40$. In contrast, for $50000/x$ where $x$ is $7$, $100$ or $7000$, we need at least a $49$-bit nonce space and at least $24$ miners.

\item \emph{Configuration C3:} For all three rate styles and all transaction values in increments of $50000$ up to $500000$, the optimal tuples are the same: $(s,r,d) = (24,49,d)$ where $36\leq d\leq 40$. 
\end{itemize}

\paragraph{Range of feasible sizes for nonce space.} We can compute and validate whether a robustly feasible tuple $(s,r,d,cost)$ has any other values $r'$ for which $(s,r',d,cost)$ is robustly feasible. For example, for all the optimal tuples $(s,r,d,cost)$ we computed above, we conclude that we may change $r$ to any $r'$ satisfying $r < r' \leq 64$.

\section{Discussion and Related Work}
\label{section:discussion}
We made Assumption~\ref{assumption:mininginvariant} only for appeal to the ROM model of the hash function used for mining. Implementations may violate this assumption, without compromising the predictive value of our models. Our Assumption~\ref{assumption:miners} is at odds with Proof of Work as used in Bitcoin. But it does simplify the reasoning about mining behavior, and makes that more akin to reasoning about Byzantine fault tolerant consensus protocols \cite{wattenhofer16}: for BFT protocols, network nodes are either honest (and so comply with protocol rules without incentives) or malicious (and so may behave in an arbitrary manner). Assumption~\ref{assumption:synchronous} is related to the assumption that a communication network be \emph{weakly synchronous}.

The mathematical model we proposed for Proof of Work did not specify details of the communication environment in which Proof of Work would operate. It would be of interest to extend our mathematical model with suitable abstractions of such a network environment, for example to reflect on upper bound on the communication delay between any two network points. This value could then be used to reflect Assumption~\ref{assumption:synchronous} in finer detail in our model. Such an extension would also allow us to investigate whether consensus protocols can be simplified by providing Proof of Work as a service with specific behavioral guarantees.

Let us discuss related work next. In \cite{Gervais16}, a quantitative framework is developed for studying the security and performance of blockchains based on Proof of Work. This framework reflects a range of parameter values such as block size and those pertaining to network propagation, and allows to determine implications of such choices on security (double-spending and selfish mining in particular) and performance. It concludes that Bitcoin could well operate at a higher transaction rate while still offering its current level of security.

In \cite{DBLP:conf/ifip11-4/Vukolic15}, the quest for the ``ultimate'' blockchain fabric is discussed: getting secure blockchains that can process high transaction volumes (performance) but do this with thousands of nodes (security). Bitcoin offers good scalability of nodes, but its transaction rate does not scale. Dually, BFT protocols \cite{lamport01,DBLP:conf/osdi/CastroL99,wattenhofer16} can offer high transaction throughput rates but their communication complexity makes use of thousands of nodes impractical. The BFT state-machine replication protocol PBFT reported in \cite{DBLP:conf/osdi/CastroL99} is designed to survive Byzantine faults in asynchronous networks~--~a proven impossibility that is circumvented with the aforementioned weak synchrony assumption in~\cite{DBLP:conf/osdi/CastroL99}. For a fixed number of $3f+1$ nodes, this resiliency to faults can be realized if at most $f$ nodes are faulty. A current leader proposes a new record to be added to the database, and three phases of communication arrive at final consensus of that addition. Views manage the transition of leadership, for example when timeouts suggest that the leader is not complying or not able to cooperate.

The cryptocurrency ByzCoin \cite{DBLP:conf/uss/Kokoris-KogiasJ16} combines ingredients from PBFT, from Bitcoin-NG (which separates leadership election and transaction verification aspects in the blockchain), and from Proof of Work to devise a hybrid blockchain: its \emph{keyblock chain} uses Proof of Work to elect the next leader, whereas the \emph{microblock} chain uses PBFT style consensus to add transactions during the current leadership. The network is open (nodes may join or leave), and the current consensus group is determined by stakes in mining that occurred within a current window of time. It uses a collective signing mechanism to reduce the communication complexity within the prepare and commit phases of the PBFT protocol.

A growing body of work uses blockchains for transactions that are not financial as such. In~\cite{DBLP:conf/sp/ZyskindNP15}, e.g., a blockchain is used as a manager for access control such that this mechanism does not require trust in a third party. The architecture of our use case can also support transactions that are not financial.

The paper \cite{DBLP:journals/corr/LundbaekDH16} discusses the work we reported in this paper in more detail. In particular, it includes a statistical validation of the random variables used in our mathematical model. In future work, we would like to support instances of our robust optimization problems in which not only $d$, $s$, and $r$ are non-constant but also other parameters of interest~--~for example the time to compute a hash $T$ or the period of time $\mu$ during which we want to avoid a conflict in the mining race. Current MINLP tools don't support such capabilities at present.

\subsection{Statistical evaluation}
We will evaluate our model by comparing its random variables,
represented in analytical form, with empirical counterparts generated
through experiments. We make this comparison in terms of both absolute
and relative error.

We ran experiments consistent with Assumptions~\ref{assumption:mininginvariant}-\ref{assumption:synchronous}
to generate data for testing the random
variables used in our model. Specifically, for a triple $(s,r,d)$ we
generated
$\lfloor 10000/s \rfloor$ mining races with hash function \emph{Double
  SHA-256} and recorded their outcome: either a
failure of all $s$ miners to proof work for level of difficulty $d$ or
an integer $rds$ saying that mining took $rds$ many rounds for proof
of work with level of difficulty $d$. The triples we studied where $(s,r,d)$ in
\begin{equation}
\label{equ:files}
\{4i\mid 1\leq i\leq 8\} \times \{2^i\mid 3\leq i\leq 7\} \times \{4,8,12,16,17,18,19,20\}
\end{equation}

\noindent The data we thus generated amount to 
\[
(5\cdot 8)\cdot \sum_{i=1}^8 \lfloor \frac{10000}{4i}\rfloor = 271720
\]
\noindent mining races.  

\paragraph{Evaluation of  $E^s(noR)$.} Suppose, for a given $(s,r,d)$, that $l$ of these $10000$ outcomes where
integers $rds$. Then the \emph{empirical failure probability} for
triple $(s,r,d)$ equals $\frac{10000-l}{10000}$.

For the evaluation of $E^s(noR)$, we compute the arithmetic average $\overline{rds}$ of all $rds$ values for a triple $(s,r,d)$ and compare this with the exact value of $E^s(noR)$ by computing the \emph{absolute} error in~(\ref{equ:absnor}) and the \emph{relative} error in~(\ref{equ:relnor})
\begin{eqnarray}
{} &{}& \size{E^s(noR) - \overline{rds}}\label{equ:absnor}\\
{} &{}& \size{E^s(noR) - \overline{rds}}/E^s(noR)\label{equ:relnor}
\end{eqnarray}

\begin{figure}
\centering
\hspace{-1.2cm}
\includegraphics[scale=0.55]{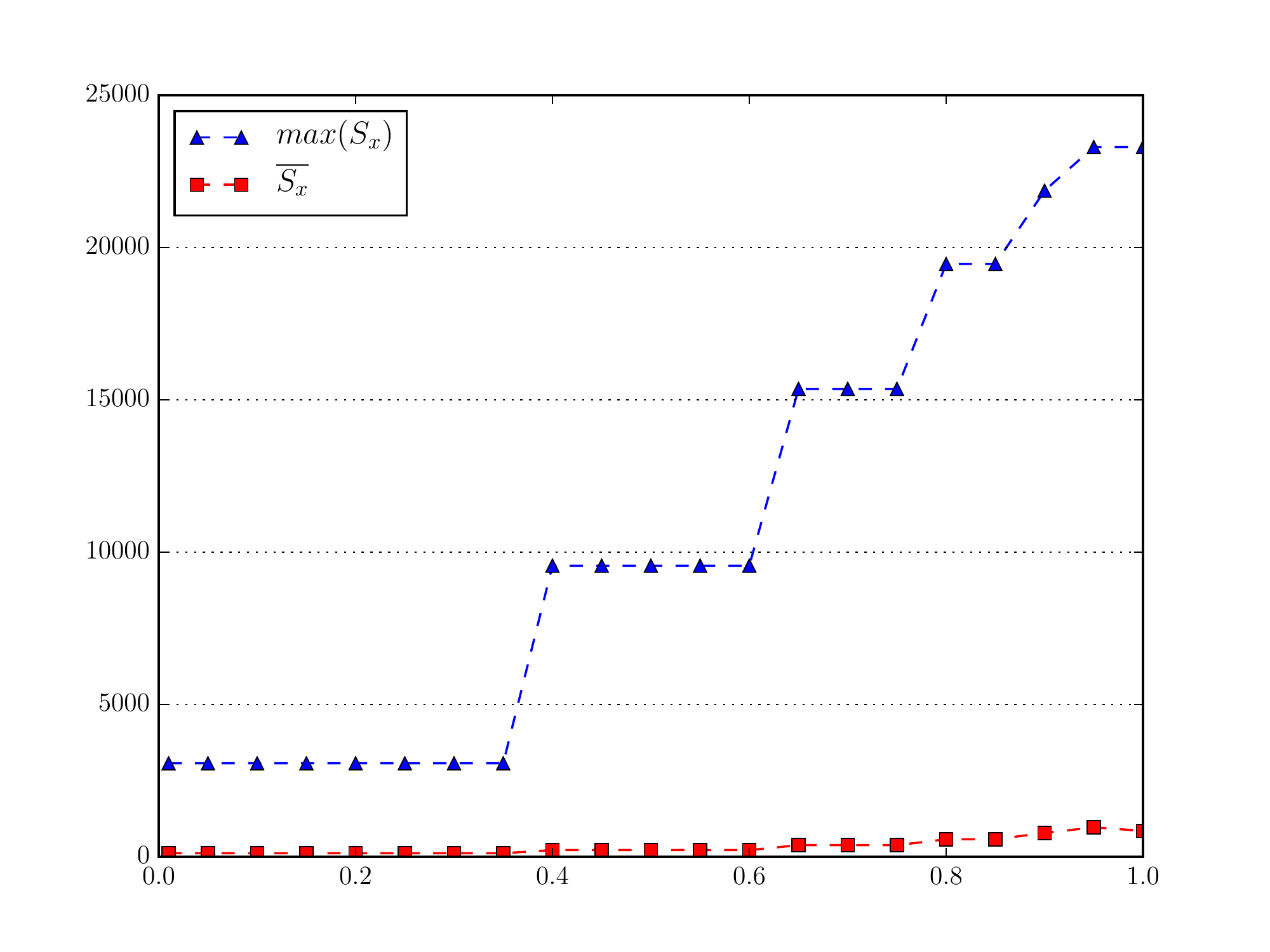}
\vspace{-0.3cm}
\caption{Discrete functions $x\mapsto \max(S_x)$ and $x\mapsto \overline{S_x}$ where $\overline{S_x}$ is arithmetic average of $S_x$ in~(\ref{equ:wx})
\label{fig:absnor}}
\end{figure}

\noindent  We analyze the absolute error first:
for each threshold value $x$ from $\{0.01+b\cdot 0.05\mid 0\leq b\leq
20\}$, let $\mathcal R_x$ be the set of all
outcomes $rds$ for all triples $(s,r,d)$ whose
empirical or analytical failure probability (i.e.\ $prob^s(\failure)$ or $\frac{10000-l}{10000}$) is strictly less than $x$. Then for the set
\begin{equation}
\label{equ:wx}
S_x= \{\size{E^s(noR) -\overline{rds}}\mid rds\in R_x\}
\end{equation}

\noindent function $x\mapsto \max(S_x)$, seen in Figure~\ref{fig:absnor}, renders a discrete graph of the maximal absolute error from set $S_x$; whereas function $x\mapsto \frac{1}{\size {S_x}}\cdot \sum_{r \in S_x} r$, named  $\overline{S_x}$ in the same figure, denotes the arithmetic average of the elements in $S_x$.
For the worst-case absolute errors, we see that they remain well below $5000$ rounds, namely at about $3070$ rounds for failure probabilities $< 0.35$. In that range of failure probabilities, the average absolute error is about $119$ rounds which is a very small difference. To illustrate,
for the time $T=0.02\cdot 10^{-9}$ to compute a sole hash used above, this would mean that the average absolute error for failure probability $< 0.35$ amounts to a time difference of
about $2.38\cdot 10^{-9}$ seconds.
We did the same analysis for the relative error in~(\ref{equ:relnor}). For example, the relative maximal error is then about $0.0323$ and the arithmetic average of relative errors is about $0.0069$.

In summary, we could show that the random variable $E^s(noR)$ accounts really well for our experimental data, and does so for failure probabilities that even extend into significant ranges such as a $30\%$ failure probability. But it also highlights that modelers need to be mindful of choosing values for $\epsilon$ in Figure~\ref{fig:firstopt} that won't invalidate the predictive value of random variables such as $E^s(noR)$. Clearly, for values of $\epsilon$ considered above, for example\ $2^{-64}$, this is a non-issue.

\paragraph{Evaluation of  $prob^s(\failure)$.} For the evaluation of $prob^s(\failure)$, the absolute
error is
\begin{equation}
\label{equ:fae}
\size{prob^s(\failure) - \frac{10000-l}{10000}}
\end{equation}

\noindent where $l$ is as above the number of successful
mining races of $10000$ overall such races for triple $(s,r,d)$ that defines $prob^s(\failure)$. The worst-case, maximal, value of all absolute errors in~(\ref{equ:fae}) ranging over all $320$ combinations of $(s,r,d)$ is about $0.0118$. The average of the absolute errors in~(\ref{equ:fae}) for all these $320$ combinations is about $0.00057$. This shows that our model of failure probabilities is very precise, both in a worst-case and in a statistical average sense when compared to our experimental data.

\paragraph{Evaluation of $prob^s(PowTime > th)$.} For the evaluation of $prob^s(PowTime > th)$, let the experiment for triple $(s,r,d)$ have been run on a machine which takes $T$ time (in seconds) to compute a sole hash. We use $T\cdot 2^d$ as a rough approximation of how long it takes to mine a block. Then we set
\(th = 1.2\cdot T\cdot 2^d\) as a reasonable test value of $th$, an
increase of twenty percent. Let $q$ be the number of times that a
reported outcome $rds$ for triple $(s,r,d)$~--~which defines $prob^s(PowTime > th)$~--~satisfies $T\cdot rds > th$. Then we want to compute $prob^s(PowTime > th)$ exactly as above, and compute the absolute error
\begin{equation}
\size{prob^s(PowTime > th) - \frac{q}{l}}
\end{equation}

\noindent as the difference between the formal and the empirical probability. However, we only consider data for tuples $(s,r,d)$ for which the constraint $\lceil (th/T) - 1\rceil < \lambda$ is met; otherwise, our computation of $prob^s(PowTime > th)$ would result in negative and therefore unsound absolute errors. This is a valid evaluation approach since
triples $(s,r,d)$ that violate $\lceil (th/T) - 1\rceil < \lambda$ would never contribute to optimizations in our models.
The worst-case, maximal, absolute error ranging over all triples $(s,r,d)$ that satisfy $\lceil (th/T) - 1\rceil < \lambda$ is about $0.003$. The arithmetic average of all these absolute errors is about $0.0001$~--~which suggests a very good fit of our model with these experimental data.

\begin{figure}
\centering
\hspace{-0.8cm}
\includegraphics[scale=0.55]{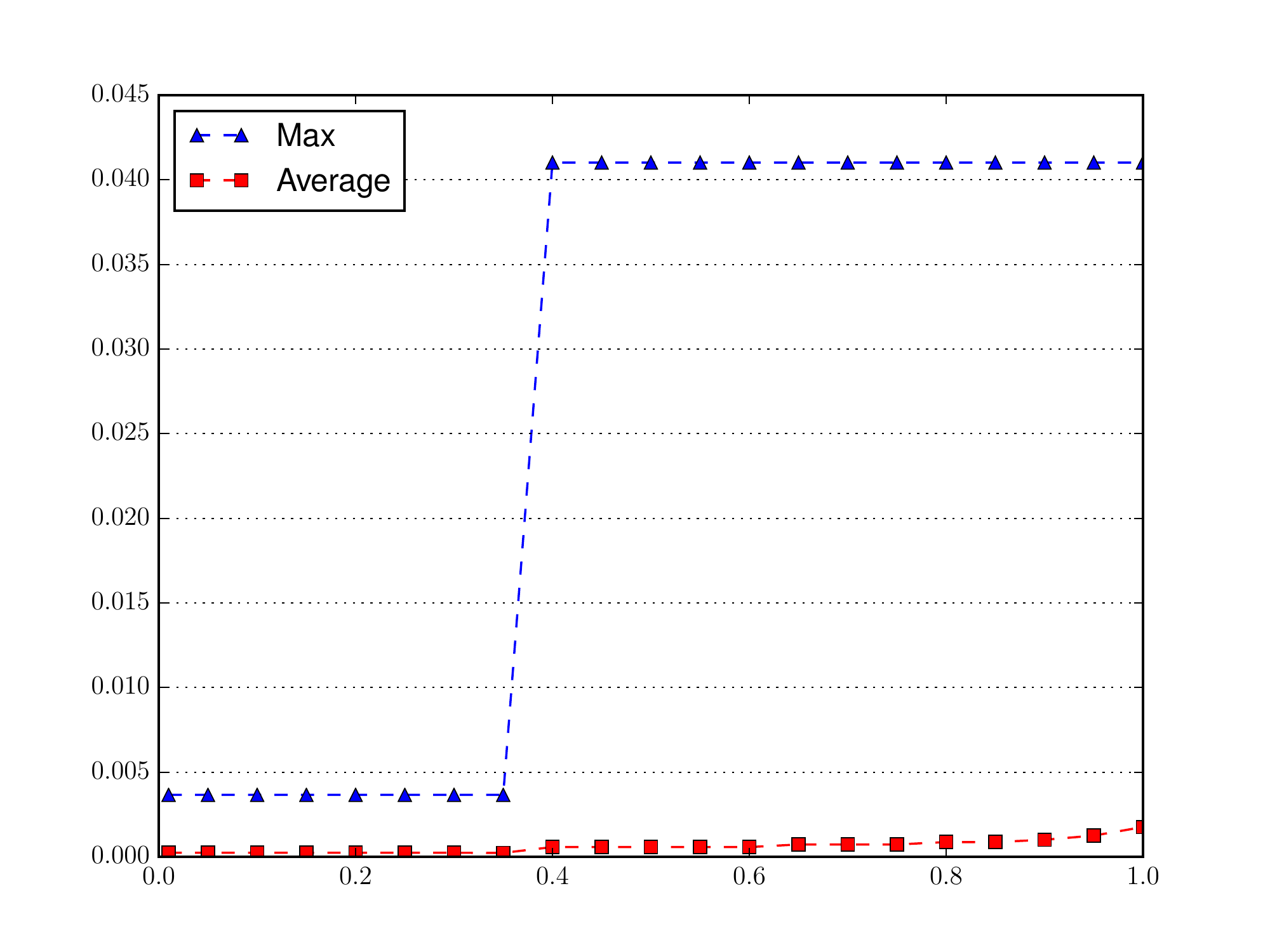}
\vspace{-0.3cm}
\caption{Graph \emph{Max} shows worst-case absolute error in~(\ref{equ:thprimeae}) for all triples $(s,r,d)$ with failure probability $< x$. Graph \emph{Average} shows the same result for arithmetic average of these absolute errors
\label{fig:th1abs}}
\end{figure}

\paragraph{Evaluation of $prob^s(PowTime < th')$.} For the evaluation of $prob^s(PowTime < th')$, we now set \(th' =
0.8\cdot T\cdot 2^d\) as a reasonable test value of $th'$, a
\emph{decrease} of twenty percent. Let $o$ be the number of times that
values $rds$ in our data for $(s,r,d)$~--~defining $prob^s(PowTime < th')$~--~satisfies $T\cdot rds < th'$. We
compute $prob^s(PowTime < th')$ as above, and report the worst-case absolute error %
\begin{equation}
\label{equ:thprimeae}
\size{prob^s(PowTime < th') - \frac{o}{l}}
\end{equation}

\noindent over all triples $(s,r,d)$
for which the empirical or analytical failure probability is strictly less
than $x$.
The results, depicted in Figure~\ref{fig:th1abs}, show that for failure probabilities below $0.35$ the \emph{worst-case} absolute errors are about $0.00366$ and so very small. The worst-case absolute errors for larger failure probabilities are about $0.041$. The arithmetic average of all these absolute errors is more variable in $x$: it ranges from $0.000235$ for $x=0.01$ to $0.00176$ (which is also the maximal arithmetic average over all $x$ considered) for $x$ equal to $1$; and these are very small values. Results for the corresponding relative error are of the same quality and so omitted.

\paragraph{Evaluation of $prob^s(disputes\ within\  \mu)$.}
For each triple $(s,r,d)$ in~(\ref{equ:files}), we 
compute the empirical probability of $prob^s(disputes\ within\ \mu)$ as follows. For each of the $k_s=\lfloor 10000/s \rfloor$ mining races, the empirical probability is $p/k_s$ where $p$ is the number of mining races in which more than one miner found proof of work for level of difficulty $d$ within $\mu$ seconds. For example, the first mining race is represented by the first $s$ entries in the file for $(s,r,d)$ and more than one miner found proof of work within $\mu$ seconds iff
more than two of these $s$ entries are both positive (meaning that $\failure$ did not occur) and less than or equal to $\lfloor \mu / T\rfloor$. This condition applies to all subsequent mining races, and so $p$ can be computed through iterated, conditioned increments of a counter.

We use this computation to generate a database of $30720 = 8\cdot 5\cdot 8\cdot (8\cdot 12)$ entries as follows. For each triple $(s,r,d)$ in~(\ref{equ:files}), we consider $8\cdot 12$ many combinations of values for $\mu$ and $T$ where
\begin{eqnarray}
{} & \mu\in \{0.02\cdot 10^{-j}\mid 1\leq j\leq 8\}\cup \{1,2,10,100\}\nonumber\\
{} & T\in \{0.02\cdot 10^{-j}\mid 1\leq j\leq 8\}
\end{eqnarray}

In particular, $\mu$ and $T$ can differ by up to ten orders of magnitude. For each such tuple $(s,r,d,\mu,T)$ as a key, its entry in the database contains the theoretical probability $prob^s(disputes\ within\ \mu)$, the empirical version of that probability, and the absolute error
\begin{equation}
\label{equ:aedispute}
{}\mid\! prob^s(disputes\ within\ \mu) - p/k_s\!\mid
\end{equation}

\noindent as discussed above. We don't consider relative errors here since the measured values are in $[0,1]$. We then query this database with queries of the form
\[
query(A,B) = \mbox{``How many entries satisfy $A$, and how many of those also satisfy $B$?''}
\]

\noindent on that database of $32720$ entries. Here, $A$ and $B$ are conditions that can be evaluated over entries, and the result of such a query is a pair $(a,b)$ where $a$ is the number of entries that satisfy $A$, and $b$ is the number of entries that satisfy $A\land B$. Subsequently, we use $\lfloor 100\cdot b/a\rfloor$ as the percentage of entries that satisfy $A$ relative to $B$.

We can use this query form to study how the match between theoretical and empirical probability changes in terms of $s$, $r$, $d$, $\mu$, and $T$.
For example, for the query
\begin{equation}
\label{equ:errorB}
query(d=x, {}\mid\! prob^s(disputes\ within\  \mu) - p/k_s\!\mid{}< {}0.1)
\end{equation}

\noindent where $x$ ranges over $\{4,8,12,16,17,18,19,20\}$, we learn the percentage of entries whose (sub)key $d$ has value $x$ and for which the absolute error is less than $0.1$. The results for that are
shown in Figure~\ref{fig:dispD}. 

We can see that empirical and theoretical probability get closer as $r$ increases, and get less close as $s$ increases,  but do so relatively slowly. There is no clear trend for that accuracy as $d$ ranges between $8$ and $20$. 
\begin{figure}
\[
\begin{array}{|c|c|c|c|c|c|c|c|c|}
\hline
d & 4 & 8 & 12 & 16 & 17 & 18 & 19 & 20\\
\hline
\% \mbox{ of $3840$ entries} & 70 & 88 & 89 & 83 & 84 & 85 & 86 & 86\\
\hline
\end{array}
\]

\[
\begin{array}{|c|c|c|c|c|c|c|c|c|}
\hline
s & 4 & 8 & 12 & 16 & 20 & 24 & 28 & 32\\
\hline
\% \mbox{ of $3840$ entries} & 88 & 87 & 83 & 83 & 83 & 82 & 82 & 82\\
\hline
\end{array}
\]

\[
\begin{array}{|c|c|c|c|c|c|c|c|c|}
\hline
r & 8 & 16 & 32 & 64 & 128\\
\hline
\% \mbox{ of $6144$ entries} & 59 & 73 & 96 & 96 & 96\\
\hline
\end{array}
\]
\caption{Percentage of entries that satisfy the query in~(\ref{equ:errorB}) as a function of $d$, $s$, and $r$ individually. These figures list the number of entries for which condition $d=x$ is true. For example, there are $6144$ entries for each value of $r$ in the database\label{fig:dispD}}
\end{figure}

Let us next understand how the absolute error behaves when $\mu$ is in $[10T,11T)$, i.e.\ when $\mu$ is at least $10\cdot T$ but smaller than $11\cdot T$.  We therefore consider queries of form 
\begin{equation}
\label{equ:errorfactortwo}
query(\lfloor \mu/ T\rfloor = 10.0, {}\mid\! prob^s(disputes\ within\  \mu) - p/k_s\!\mid{}< {}x)
\end{equation}

\noindent where $x$ is in $\{10^{-j}\mid 2\leq j\leq 10\}$. 
There are $2240$ entries in the database for which $\lfloor \mu/ T\rfloor = 10.0$ holds, i.e.\ for which $10T\leq \mu < 11T$ holds. For $x=0.01$, $98\%$ of these entries have  absolute error less than $x$. This decreases to $86\%$ for $x=0.001$, with further decreases to $75\%$, $67\%$, and $62\%$ for $x$ being $0.0001$, $0.00001$, and $0.000001$ (respectively). The percentage stays at $42\%$ for $x$ being $0.0000001$, $0.00000001$, and $0.000000001$. For the modified queries
\[
query((\lfloor \mu/ T\rfloor = 10.0)\land prob^s(\failure) < 0.001), {}\mid\! prob^s(disputes\ within\  \mu) - p/k_s\!\mid{}< {}x)
\]

\noindent the results are similar but slightly worse, $(97\%, 83\%,70\%,59\%, 55\%, 37\%,37\%,37\%,37\%)$ for descending values of $x$.
But these results suggest robustness of our statistical measure, since these results allow for variability in $s$, $d$, and $r$.
This robustness is corroborated by 
\begin{equation}
query(prob^s(\failure) < 0.1), {}\mid\! prob^s(disputes\ within\  \mu) - p/k_s\!\mid{}< {}x)
\end{equation}

\noindent being $(18926, 21504)$ and so $88\%$ of all entries with failure probability below $0.1$ have an absolute error smaller than $0.01$. This percentage decreases to $84\%$ when the absolute error is smaller than $0.005$. 

Finally, let us report how the absolute error evolves when the ratio $\lfloor \mu / T\rfloor$ increases. Consider the query
\begin{equation}
\label{equ:muTratio}
query(\lfloor \mu / T\rfloor = x, {}\mid\! prob^s(disputes\ within\  \mu) - p/k_s\!\mid{}< 0.1)
\end{equation}

\noindent where $x$ varies as seen in Figure~\ref{fig:muTratio}. This shows that the absolute error is less than $0.1$ in most cases even when $\mu$ is orders of magnitude larger than $T$.
\begin{figure}
\[
\begin{array}{|c|c|c|c|c|c|c|c|c|c|c|c|c|}
\hline
x & 0 & 1 & 10 & 99 & 100 & 500 & 999 &1000 & 5000 & 10000 & 49999 & 50000\\
\hline
\% \mbox{ of entries} & 90 & 89 & 98 & 96 & 96 & 95 & 95 & 95 & 88 & 83 & 71 & 71\\
\hline
\end{array}
\]

\caption{Percentage of entries that satisfy the query in~(\ref{equ:muTratio}) as a function of $x$.
The percentage equals $70\%$ for all $x = 99999$ and stays at that percentage for those values up to $x=5\cdot 10^{11}$ that may change the percentage given the particular keys in the database\label{fig:muTratio}}
\end{figure}

\paragraph{Summary of Evaluation.} These findings above are evidence of the validity of our random
variables and their use in our modelling approach. But they also highlight that
failure
probabilities of mining larger than 30\% may require caution in using
our approach. Actual implementations would not want to realize such 
large failure probabilities in their design phase. We emphasize that these
experiments indirectly depended on security properties of the underlying
hash function.

\section{Conclusions}
\label{section:conclusion}
In this paper we considered blockchains as a well known mechanism for the creation of trustworthiness in transactions, as pioneered in the Bitcoin system \cite{nakamoto08}. We studied how blockchains, and the choice and operation of cryptographic puzzles that drive the creation of new blocks, could be controlled and owned by one or more organizations. Our proposal for such governed and more central control is that puzzle solvers are mere resources procured by those who control or own the blockchain, and that the solution of puzzles does not provide any monetary or other reward. In particular, a newly solved block will not create units of some cryptocurrency and there is therefore no inherent incentive in solving puzzles.

The absence of incentives thus avoids the well known problems with game-theoretic behavior, for example that seen within and across mining pools in Proof of Work systems such as Bitcoin.
Furthermore, it lends itself well to the development of proprietary or private blockchains that are \emph{domain-specific} and whose specific purpose may determine who can access it and in what ways. We illustrated this idea with a use case in which financial transactions recorded within conventional accounts would be recorded as hashes within a governed blockchain and where it would be impractical to use hash chains, due to non-linearizability of transaction flows, and due to their malicious manipulability to pass auditory requirements.

A blockchain design should of course have specifications of its desired behavior, including but not limited to
the expected time for creating a new block, resiliency against corruption of some of the puzzle solvers, service level guarantees such as a negligible probability that the time needed for creating a new block exceeds a critical threshold or a negligible probability that more than one solver does solve a puzzle within a specified period of time.

We developed mathematical foundations for specifying and validating a crucial part of a governed blockchain system, the solving of cryptographic puzzles~--~where we focussed on Proof of Work. In our approach, owners of a blockchain system can specify allowed ranges for the size of the shared nonce space, the desired level of difficulty, and the number of miners used; and they can add mathematical constraints that specify requirements on availability, security, resiliency, and cost containment~--~such as the ones just discussed. This gives rise to MINLP optimization problems that we were able to express in analytical form, by appeal to the ROM model of cryptographic hash functions used for cryptographic puzzles.

We then wrote an algorithm that can solve such MINLP problems for sizes of practical relevance.
We illustrated this on some instances of that MINLP problem. This demonstrated that we have the capability of computing optimal design decisions for a governed Proof of Work system, where resiliency is modeled through robust optimization. This \emph{mining calculus} also supports change management. For example, if we wanted to increase mining capacity and/or mining resiliency, our mathematical model could be used to determine how many new miners are needed to realize this~--~be it for the same or better hardware specifications. For another example, our tool could be used to determine optimal numbers of used miners or parameters by reacting to new energy prices.

Our approach and mathematical model are consistent with the consideration of several
organizations controlling and procuring heterogeneous system resources, with each such organization having its bespoke blockchain, and with the provision of puzzle solving as an outsourced service. We leave the refinement of our mathematical models to such settings as future work. It will also be of interest to develop mathematical techniques for the real-time analysis of such blockchains, for example, to assess statistically whether the observed history of cryptographic puzzle solutions is consistent with the design specifications.

We hope that the work reported in this paper will provoke more thinking about the design, implementation, and validation of blockchains that are centrally~--~or in a federated manner~--~owned and controlled and that
may fulfill domain-specific needs for the creation of trustworthiness. We believe that many domains have such needs that the approach advocated in this paper might well be able to meet:
existing financial processes and payment workflows (which conventional cryptocurrencies are more likely to replace than to adequately support), trustworthiness of information in Internet of Things systems (where the difficulty of the puzzle, for example, may have to be contained), but also systems that have governed blockchains at the heart of their initial design (for example, a payment system in which the temporal and causal history of payments, logs, and audits is recorded on the blockchain).

\bigskip
\noindent {\bf Acknowledgements:} This work was supported by the UK Engineering and Physical Sciences Research
Council  with a Doctoral Training Fees Award for the first author and with projects [grant numbers EP/N020030/1 and EP/N023242/1]. We expressly thank Ruth Misener for having run some of our models on state-of-the-art global MINLP solvers~--~the tools ANTIGONE
\cite{misener-floudas:2014}, BARON \cite{tawarmalani-sahinidis:2005}, and
SCIP \cite{vigerske:2012}~--~to conclude that these solvers judge the models to be infeasible, although they are feasible.\footnote{In fairness, these global MINLP solvers were not built for dealing with numerical problems such as those encountered in our models.}

\end{document}